\documentclass [a4paper,12pt,epsfig]{article}
\usepackage{latexsym}
\usepackage{graphicx}

\begin{document}

\begin{center}
{\large \bf Ion Irradiation Induced Effects in Metal Nanostructures}\\
\end{center}

\begin{center}
B. Satpati, P. V. Satyam, T. Som and B. N. Dev$^{\ast}$\\

\vspace{0.2in}
Institute of Physics, Sachivalaya Marg, Bhubaneswar 751 005, India\\
\end{center}

\vspace{0.2in}
\begin{center}
\textbf{ABSTRACT}
\end{center}

High resolution transmission electron microscopy (HRTEM)  and Rutherford
backscattering spectrometry (RBS) are used to study the ion induced
effects in Au, Ag nanostructures grown on Si and thermally grown SiO$_2$
substrates. Au and Ag films ($\sim$ 2 nm) are prepared by thermal
evaporation under high vacuum condition at room temperature (RT). These
films were irradiated with MeV Au ions also at RT. Very thin films of Au
and Ag deposited on silicon substrates (with native oxide) form isolated
nano-island structures due to the non-wetting nature of Au and Ag. Ion
irradiation causes embedding of these nanoislands into the substrate.  
For Ag nanoislands with diameter 15 - 45 nm, the depth of the embedding
increases with ion fluence and the nano particles are fully submerged into
Si and SiO$_2$ substrate at a fluence of $\sim$ 5$\times$10$^{14}$ ions
cm$^{-2}$ without any mixing.  Au nanoparticles (diameter 6 - 20 nm), upon
ion irradiation, forms embedded gold-silicide in the case of Si substrate
and show lack of mixing and silicide formation in the case of SiO$_2$
substrate system.\\


\vspace{2.0in}

\noindent{$^{\ast}$ E-mail:
bhupen@iopb.res.in, Fax: +916742300142.}\\

\newpage
\section {Introduction}

As microelectronic devices and interconnects are reduced in size and
electrical connections are made using ultra-low resistivity materials,
many applications exist for noble metal thin films.  On the other hand,
energetic ion beams have been exploited in different ways in the field of
materials science.  The ion beam mixing is widely used for generating new
phases specially silicides with the help of low energy ion beams.  These
phases have many potential applications in modern semiconductor industry
for making contacts, interconnects, Schottky barriers, insulating layers,
protective coatings, etc. \cite{Nastasi-1, Nastasi-2, Aboelfotoh}.
Consequently ion beam modification of noble metal thin films has been an
active area of research. Recently, this area has extended into
modification of metal nanoparticles.\\
      
\noindent Energetic ion beams have been utilized in synthesizing and
modifying nanostructures \cite{Jacobsohn, White, Joseph}. The application
of ion beams to modify properties of nanomaterials has now become an
important topic due to the emerging technological importance. As
ion-irradiation is an athermal process, properties of nanomaterials could
be modified where such modifications are not feasible by conventional
methods \cite {Dillen}. The behavior of nanoparticles on surfaces is
important for applications in catalysis, nanomagnetics, and
optoelectronics.  In most cases, nanoparticles are not thermodynamically
stable on surfaces and would undergo smoothing reactions were it not for
kinetic constraints. Nanoparticles that are immiscible with the substrate
can undergo smoothing reactions by two mechanisms, wetting or burrowing
\cite{Zimmermann}. When nanoparticles have surface energies that are
sufficiently smaller than that of the substrate, they spread over the
surface. When the surface energy is large, the nanoparticles tend to
burrow. Ion-beam irradiation of Pt nanoparticles supported on a SiO$_2$
substrate leads to the burrowing of the particles into the substrate
\cite{Hu}. In other related studies on ion-induced effects in nanoislands,
Baranov et al. \cite{Baranov} reported sputtering from nanoclusters with
cluster ion beams. Among other effects induced by ion irradiation,
sputtering of target atoms and modifications of target surface and
interfaces are two important aspects. A radiation damage model that
explains the production of interstitial atoms, vacant lattice sites, and
displacement spikes was introduced several decades ago to understand
irradiation effects in materials \cite{Brinkman}. For ion-induced effects
in target materials, the composition and morphological variations due to
thermal spikes were investigated both experimentally \cite {Merkle,
Birtcher, Farenzena, Satyam} and theoretically \cite {Nordlund, Averback,
Bringa, Yang, Rubia}. In our previous studies, we have reported about a
push-in effect for silver nanoislands deposited on silicon substrates in
comparison with gold nanoislands on silicon under MeV ion irradiation
\cite{Satpati-1} and absence of push-in effect in case of thick continuous
Au film \cite{Satpati-2}. About 2 nm thick native oxide was present on the
silicon substrate prior to film deposition.  When a thin layer of SiO$_2$
is present on Si, the barrier formed at the metal SiO$_2$ interface is
much higher than that obtained directly at the metal-semiconductor contact
because of the large band gap of SiO$_2$. Here, we present results about
push-in effect for gold and silver nanoislands deposited on silicon
substrates with native oxide in comparison with gold and silver
nanoislands deposited on a thermally grown SiO$_2$ substrate.\\

\noindent Ag and Au nanoparticles were selected for these experiments
since noble metal nanoparticles on amorphous substrates such a s SiO$_2$,
have been widely used for heterogeneous catalysis.  Moreover, Au and Ag
are relatively inert in air, making possible {\it ex-situ}
characterization of irradiated samples.

\section{Experimental} 

Si wafers with native oxide ($\sim$ 2 nm) and Si wafers with a thick
($\sim$ 280 nm) thermally grown SiO$_2$ layer were successively cleaned
with trichloroethylene, acetone, and methanol in an ultrasonic bath and
introduced into the thin film deposition chamber. Au (1.3 nm) and Ag (1.2
nm) films were deposited separately on such substrates by thermal
evaporation at a rate of 0.1 nm/s. In the case of Au deposition was also
carried out on HF-etched Si substrates. The chamber pressure during
deposition was 8$\times$10$^{-6}$ mbar. Irradiation was carried out with
1.5 MeV Au$^{2+}$ ions from the Peletron accelerator in our institute at
an impact angle of 60$^{\circ}$ between the sample surface-normal and the
incident ion beam and with fluences ranging from 5$\times$10$^{13}$ to
5$\times$10$^{14}$ ions cm$^{-2}$. The ion current for the irradiations
was $\approx$ 30 nA and the ion beam was rastered over an area of
1$\times1$ cm$^2$. During irradiation, 2$\times$10$^{-7}$ mbar pressure
was maintained in the irradiation chamber. Transmission electron
microscopy (TEM) measurements were carried out with 200 keV electrons
(JEOL JEM-2010). The cross-section and the planar samples were prepared
using mechanical thinning followed by 3.0 keV Ar ion milling using a GATAN
PIPS equipment. The amount of deposited material (the effective film
thickness)  was measured by Rutherford backscattering spectrometry using 2
MeV He$^{2+}$ ions.

\section{Results and Discussions}

Gold films of effective thickness 1.3 $\pm$ 0.1 nm on
native-oxide/Si(100), HF-etched Si(100) and thermally grown
SiO$_2$/Si(100) (denoted later as SiO$_2$/Si(100)) and silver films of
thickness 1.2 $\pm$ 0.1 nm on native-oxide/Si (100) and SiO$_2$/Si(100)
have been used for the present study. Gold and silver grow as nanoislands
rather than uniform films. The amount of deposited material is expressed
in terms of an effective thickness, which would be the actual film
thickness if the film were deposited as a film of uniform thickness.  The
maximum height of the Au and Ag islands is $\sim$ 30 nm. The range of the
1.5 MeV gold ions in Si at an impact angle of 60$^{\circ}$ is found to be
$\approx$ 310 nm using the SRIM 2003 range calculation \cite {Biersack};
the ranges of these ions in Au and Ag are $\approx$ 90 nm and $\approx$
120 nm respectively.  This means that ions, after interacting with the
metal islands, get buried into the substrate.

\subsection{Au/native-oxide/Si}
      
TEM results from Au-deposited samples are shown in Fig. 1. Fig. 1(a)  and
(b) show plan-view TEM micrographs of an as-deposited and an
ion-irradiated (fluence: 1$\times$10$^{14}$ ions cm$^{-2}$) sample
respectively. Gold islands are isolated with average particle size
(diameter) 11.1 $\pm$ 0.1 nm, a standard deviation ($\sigma$) of 5.1 $\pm$
0.3 nm and 26\% surface coverage of islands. These values are obtained
using several TEM micrographs and ImageJ analysis package. Fig. 1(c) shows
a cross-sectional TEM (XTEM) micrograph of the sample shown in Fig. 1(b).
From XTEM micrographs of an as-deposited sample, Au-island thickness
(height) is found to vary mostly between 6 to 12 nm (data not shown). It
is evident from Fig. 1(b) and 1(c) that craters are formed in Au
nanoislands and Au has been embedded into the silicon substrate upon ion
bombardment.\\
      
\noindent In order to understand more about the mixing and phase
formation, we have carried out high resolution lattice imaging from the
regions displaying buried structures. Fig. 1(d) shows such a high
resolution XTEM (HR-XTEM) micrograph of a mixed region of Fig. 1(c) marked
by a rectangle. The lattice image shows a d-spacing of 0.293 $\pm$ 0.01
nm. This value does not match with any of the inter-planar spacing in pure
gold [19]. This value is very close to the (210) interplaner spacing of
Au$_5$Si ($d_{210}$ for Au$_5$Si is 0.297 nm \cite {Suryanarayana}).
Results from RBS measurements are shown in Fig. 2, which presents spectra
taken form pristine, ion-irradiated and aqua-regia-treated gold
nanoislands on native-oxide/silicon substrate. We analyze these spectra
based on the previous knowledge that aqua-regia removes Au by etching,
while it is unable to remove gold-silicide. The RBS spectrum from
aqua-regia- treated pristine sample native-oxide/Au/Si shows the absence
of gold signal, which is a natural consequence of etching of deposited Au
by aqua-regia. RBS measurements on an aqua-regia-treated sample, following
the irradiation with a fluence of 1$\times$10$^{14}$ cm$^{-2}$, show the
presence of almost as strong Au signal as it was before etching. As gold
silicide cannot be removed by aqua-regia treatment, this indicates that at
this fluence almost all Au form a reacted gold silicide either on the
surface or within the silicon substrate. Considering these results
together with the XTEM micrograph (Fig. 1(c)), it is clear that this
gold-silicide is predominantly inside the silicon substrate. That is, ion
bombardment has caused embedding of Au nanoparticles into Si and a
concomitant reaction forming gold silicide.

\subsection{Au/HF-etched-Si}

Fig. 3(a) shows an XTEM micrograph of Au/HF-etched-Si substrate upon
irradiation with an ion fluence of 1$\times$10$^{14}$ cm$^{-2}$. Form the
micrograph it is evident that a band ($\sim$ 20 nm) of reacted material
forms at the sub-surface region. It should be noted that when Au deposited
on a clean (without native oxide) Si substrate it wets more (data not
shown) than that of Au deposited on native oxide layer atop the Si
substrate. This may be understood from their respective surface free
energies. Fig. 3(b) shows a high resolution XTEM micrograph of the region
of Fig. 3(a) marked by a rectangle. The lattice image shows a d-spacing of
0.293 $\pm$ 0.01 nm. This value does not match with any of the
inter-planar spacing available for the pure gold. This value again is very
close to the (210) interplaner spacing of Au$_5$Si ($d_{210}$ for Au$_5$Si
is 0.297 nm \cite {Suryanarayana}).

\subsection{Au/SiO$_2$/Si}

For the Au nanoparticles grown on a thermally grown SiO$_2$, ion
irradiation shows only embedding but no formation of gold-silicide. Fig.
3(c) shows an XTEM micrograph of ion- irradiated Au/SiO$_2$/Si (fluence:
1$\times$10$^{14}$ cm$^{-2}$). Form Fig. 3(c) it is evident that Au
nanoislands are partially submerged into SiO$_2$ substrate. We perform
lattice imaging from embedded region. One such image is shown in Fig.
3(d). The lattice image shows a d-spacing of 0.191 $\pm$ 0.01 nm. This
value is very close to the (200) interplaner spacing of bulk Au ($d_{200}$
for Au is 0.203 nm). Inset in Fig. 3(d) showing FFT corresponding to
region marked by a rectangle, displaying a [011] fcc diffraction pattern.
These results confirm that the embedded part of the nanoparticle is
unreacted Au.

\subsection{Ag/native-oxide/Si}

Fig. 4(a) shows XTEM micrograph of an as-deposited sample. From plan-view
TEM images, the average particle size (diameter) for the Ag islands is
found to be 31.6 $\pm$ 0.3 nm with $\sigma$ = 14.3 $\pm$ 1.0 nm and with
22 \% area coverage (data not shown). From XTEM micrographs, Ag-island
height is found to vary between 10 to 30 nm. Fig. 4(b), (c) and (d) show
XTEM micrographs of ion-irradiated Ag/native-oxide/Si with ion fluence of
5$\times$ 10$^{13}$, 1$\times$10$^{14}$ and 5$\times$10$^{14}$ cm$^{-2}$
respectively. Form Figs. 4(b)-(d) it is clear that depth of embedding
increases with ion fluence and at a fluence of $\sim$ ions
5$\times$10$^{14}$ cm$^{-2}$, the Ag nanoparticles are fully submerged
into the Si substrate without any reaction. Fig. 5(a) shows an HR-XTEM
micrograph of embedded Ag particle inside the Si matrix after irradiation
with ion fluence of 5$\times$10$^{14}$ cm$^{-2}$. The lattice image shows
a d-spacing of 0.223 $\pm$ 0.01 nm, which is close to (111) interplanar
spacing of bulk Ag ($d_{111}$ for Ag is 0.236 nm). Fig. 5(b), showing FFT
corresponding to region marked by a rectangle in Fig. 5(a), displays a
[011] fcc diffraction pattern.

\subsection{Ag/SiO$_2$/Si}

Fig. 5(c) shows an XTEM micrograph of Ag nanoparticles on SiO$_2$ after
irradiation with ion fluence of $\sim$ ions 5$\times$10$^{14}$ cm$^{-2}$.
In this case also Ag nanoislands are fully submerged into SiO$_2$
substrate without any reaction. The HR-XTEM micrograph in Fig. 5(d) shows
an embedded Ag particle inside SiO$_2$ matrix (magnified portion of Fig.
5(c) marked by a rectangle). The lattice image shows a d-spacing of 0.212
$\pm$ 0.01 nm, which is close to (200) inter planar spacing of bulk Ag
($d_{200}$ for Ag is 0.204 nm). The lattice spacing for Ag nanoparticles
inside the Si or SiO$_2$ matrix could be somewhat different than bulk Ag
lattice spacing.

\section{Conclusions}

We have shown that nanoparticles of Au and Ag on native-oxide-Si and on
SiO$_2$/Si can be embedded into the substrates by ion bombardment. Not
only just embedding, there is also a concomitant formation of a reacted
material (gold-silicide) in case of Au/native-oxide-Si. The mixed phase of
Au-Si is found to be crystalline in nature. No mixed phase formed in case
of Au/SiO$_2$/Si, Ag/native-oxide-Si and Ag/SiO$_2$/Si. The high eutectic
temperature, high and positive heat of mixing of Ag-Si system compared to
the Au-Si system could be responsible for the lack of mixing in the Ag-Si
system. If Co nanoparticles can be embedded into Si and an embedded
nanoscale cobalt silicide can be formed, it would be a prospective
candidate for nanotransistors. As a Schottky barrier can be formed at the
cobalt-silicide/silicon interface, depositing Si on this cobalt-silicide,
in principle, a nanoscale metal base transistor can be fabricated.

\newpage

Figure captions:

\begin{figure}[htbp]
\caption{\label{Fig1} (a) A plan-view TEM image from an as-deposited film
showing Au nanoparticles on a native-oxide/Si substrate (b) Plan-view TEM
image of Au nanoparticles on a native-oxide/Si substrate following 1.5 MeV
Au$^{2+}$ ion bombardment with a fluence of 1$\times$10$^{14}$ ions
cm$^{-2}$ at $60^{\circ}$ impact angle; (c) An XTEM image of the sample in
(b); (d) High resolution image from the region marked by a rectangle in
(c).} \end{figure}

\begin{figure}[htbp]
\caption{\label{Fig2} RBS spectra form as-deposited, ion-irradiated and
aqua-regia treated 1.3 nm Au/native-oxide/Si samples.  Results confirm
embedding of Au with concomitant silicide formation.} \end{figure}

\begin{figure}[htbp]
\caption{\label{Fig3}Cross-sectional TEM images of Au nanoparticles on
HF-etched Si substrate and on thermally grown SiO$^2$/Si substrate
irradiated with 1.5 MeV Au$^{2+}$ at 60$^{\circ}$ impact angle: (a) Au
nanoparticles on HF-etched Si substrate after irradiation with ion fluence
of 1$\times$10$^{14}$ ions cm$^{-2}$, (b) High resolution image from
region marked by a rectangle in (a), (c) Au nanoparticles on SiO$^2$/Si
substrate after irradiation with fluence of 1$\times$10$^{14}$ ions
cm$^{-2}$, (d) High resolution image of one embedded particle in (c).
Inset in Fig. (d) Showing FFT corresponding to the region marked by a
rectangle, displaying a [011] fcc diffraction pattern.} \end{figure}

\begin{figure}[htbp]
\caption{\label{Fig4}Cross-sectional TEM images of Ag nanoparticles on Si
substrate irradiated with 1.5 MeV Au$^{2+}$ at 60$^{\circ}$ impact angle:
(a) before irradiation, (b) after irradiation with ion fluence of
5$\times$10$^{13}$ cm$^{-2}$, (c) ion fluence of 1$\times$10$^{14}$
cm$^{-2}$, (d) ion fluence of 5$\times$10$^{14}$ cm$^{-2}$.} \end{figure}

\begin{figure}[htbp]
\caption{\label{Fig5}(a) HR-XTEM images of Ag nanoparticles inside Si
matrix after irradiation with 1.5 MeV Au$^{2+}$ at 60$^{\circ}$ impact
angle, (b) showing FFT corresponding to region marked by a rectangle in
(a), displaying a [011] fcc diffraction patter, (c) Ag nanoparticles on
SiO$_2$ after irradiation with ion fluence of 5$\times$10$^{14}$
cm$^{-2}$, (d) HR-XTEM images of Ag nanoparticles inside SiO$_2$ matrix
from a region marked by a rectangle in (c). The solid line drawn in (a)
and (d) represents the surface.} \end{figure}


\begin{thebibliography}{99}

\bibitem{Nastasi-1}M. Nastasi, J.W. Mayer, and J.K. Hirvonen, {\it
``Ion-Solid Interaction: Fundamentals and Applications''}, (Cambridge
University Press, 1996).

\bibitem{Nastasi-2} M. Nastasi and J.W. Mayer, {\it ``Ion Beam Mixing in
Metallic and Semiconductor Materials''}, Materials Science and
Engineering, R12, No. 1, May 1994

\bibitem{Aboelfotoh} M.O. Aboelfotoh, C.L. Lin, and J.M. Woodall: Appl.
Phys. Lett. {\bf 65}, (1994) 3245.

\bibitem{Jacobsohn} L.G. Jacobsohn, M.E. Hawley, D.W. Cooke, M.F. Hundley,
J.D. Thompson, R.K. Schulze, and M. Nastasi: J. Appl. Phys. {\bf 96}, (2004)
4444.

\bibitem{White} C.W. White, S.P. Withrow, J.M. Williams, J.D. Budai, A.
Meldrum, K.D. Sorge, J.R. Thompson, and L.A. Boatner: J. Appl. Phys. {\bf 95},
(2004) 8160.

\bibitem{Joseph} B. Joseph, S. Mohapatra, B. Satpati, P.K. Kuiri, H.P.
Lenka and D.P. Mahapatra: Nucl. Instr. Meth. Phys. Res. B (in press).

\bibitem{Dillen} T. van Dillen, A. Polman, W. Fukarek, A. van Blaaderen:
Appl. Phys. Lett. {\bf 78}, (2001) 910.

\bibitem{Zimmermann} C.G. Zimmermann, M. Yeadon, K. Nordlund, J.M. Gibson,
and R.S. Averback: Phys. Rev. Lett. {\bf 83}, (1999) 1163.

\bibitem{Hu} Xiaoyuan Hu, David G. Cahill, and R.S. Averback: J. Appl.
Phys. {\bf 92}, (2002) 3995.

\bibitem{Baranov} I. Baranov, A. Brunelle, S. Della-Negra, D. Jacquet, S.
Kirillov, Y.L. Beyee, A. Novikov, V. Obnorskii, A. Pchelintsev, K. Wien,
S. Yarmiychuk: Nucl. Instrum. Methods Phys. Res. B {\bf 193}, (2002) 809.

\bibitem{Brinkman} J.A. Brinkman: Am. J. Phys. {\bf 24}, (1956) 246.

\bibitem{Merkle} K.L. Merkle, W. Jager: Philos. Mag. A  {\bf 44}, (1981) 741.

\bibitem{Birtcher} R.C. Birtcher, S.E. Donnelly, S. Schlutig: Phys. Rev.
Lett. {\bf 85}, (2000) 4968, and references therein.

\bibitem{Farenzena} L.S. Farenzena, R.P. Livi, M.A. de Araujo, G. Garcia
Bermudez, R.M. Papaleo: Phys. Rev. B {\bf 63}, (2001) 104108, and references
therein.

\bibitem{Satyam} P.V. Satyam, J. Kamila, S. Mohapatra, B. Satpati, D.K.
Goswami, B.N. Dev, R.E. Cook, L. Assou.d, J. Wang: J. Appl. Phys. {\bf 93},
(2003) 6399.

\bibitem{Nordlund} K. Nordlund, J. Keinonen, M. Ghaly, R.S. Averback:
Nature {\bf 398}, (1999) 49.

\bibitem{Averback} R.S. Averback: Phys. Rev. Lett. {\bf 72}, (1994) 364.

\bibitem{Bringa} E.M. Bringa, K. Nordlund, J. Keinonen: Phys. Rev. B {\bf 64},
(2001) 235426.

\bibitem{Yang} Q. Yang, T. Li, B.V. King, R.J. MacDonald: Phys. Rev. B 
{\bf 53}, (1996) 3032.

\bibitem{Rubia} T. Diaz de la Rubia, R.S. Averback, R. Benedek, W.E. King:
Phys. Rev. Lett. {\bf 59}, (1987) 1930.

\bibitem{Satpati-1} B. Satpati, P. V. Satyam, T. Som and B. N. Dev: Appl.  
Phys. A, {\bf 79}, (2004) 447.

\bibitem{Satpati-2} B. Satpati, P. V. Satyam, T. Som and B. N. Dev: J.
Appl. Phys. {\bf 96}, (2004) 5212.

\bibitem{Biersack} J.P. Biersack and L.G. Haggmark: Nucl. Instrum. Meth.,
{\bf 174}, (1980) 257; http://www.srim.org/

\bibitem{Suryanarayana} Suryanarayana et al.,: Mater. Sci. Eng. {\bf 13} (1974)
73.

\end{thebibliography}
\end{document}